\documentclass{article}

\usepackage{arxiv}

\usepackage[utf8]{inputenc} 
\usepackage[T1]{fontenc}    
\usepackage{hyperref}       
\usepackage{url}            
\usepackage{booktabs}       
\usepackage{amsfonts}       
\usepackage{nicefrac}       
\usepackage{microtype}      
\usepackage{lipsum}
\usepackage{graphicx}
\usepackage{multirow}
\usepackage{amsmath}
\graphicspath{ {./images/} }

\title{The Role of Task Complexity in Reducing AI Plagiarism: A Study of Generative AI Tools}

\author{
 Sacip Toker \\
  Information Systems Engineering\\
  Atılım University\\
  Ankara, Türkiye \\
  \texttt{sacip.toker@atilim.edu.tr} \\
   \And
 Mahir Akgun \\
  College of Information Sciences and Technology\\
  Pennsylvania State University\\
  University Park, PA 16827 \\
  \texttt{makgun@psu.edu} 
  \\
}

\begin{document}
\maketitle
\begin{abstract}
This study investigates whether assessments fostering higher-order thinking skills can reduce plagiarism involving generative AI tools. Participants completed three tasks of varying complexity in four groups: control, e-textbook, Google, and ChatGPT. Findings show that AI plagiarism decreases as task complexity increases, with higher-order tasks resulting in lower similarity scores and AI plagiarism percentages. The study also highlights the distinction between similarity scores and AI plagiarism, recommending both for effective plagiarism detection. Results suggest that assessments promoting higher-order thinking are a viable strategy for minimizing AI-driven plagiarism.
\end{abstract}


\section{Introduction}
Students in higher education are required to produce various forms of text, such as assignments, reports, essays, and theses, making writing one of their most frequent activities. However, producing these outcomes is often challenging, especially for students with poor language skills \cite{hu2015chinese, ison2018empirical}. These difficulties can lead to academic misconduct, including plagiarism \cite{jereb2018factors}. Plagiarism ranges from minor issues like paraphrasing without citation to more severe forms, such as using substantial portions of another's work without crediting the source \cite{larkham2002plagiarism, bennett2005factors}. Irrespective of the form of plagiarism, while some countries have seen sharp increases in plagiarism, others have remained stable, showing that the problem persists globally. For instance, contract cheating, where students pay others to complete work, is rising worldwide, affecting millions \cite{newton2018common}. In Austria, around 22\% of students admitted to plagiarism when anonymity was assured \cite{hopp2021prevalent}. Similarly, Australia saw no improvement in plagiarism rates after 2014, despite continued efforts \cite{curtis2021plagiarism}. These findings highlight the importance of sustained efforts to prevent plagiarism, which turns attention to the factors contributing to plagiarism.

Plagiarism is influenced by individual (e.g., unfamiliarity with the concept), academic (e.g., poor writing skills), curriculum (e.g., excessive workload), and technological (e.g., paper mills and developments in ICT) factors. For example, raising awareness of plagiarism ethics can significantly improve students' ethical judgment \cite{prashar2023plagiarism}. Conversely, poor understanding of plagiarism increases its prevalence \cite{maxwell2008does}. However, even when students claim to understand plagiarism, many still engage in it, both intentionally and unintentionally, primarily due to pressures for high grades, laziness, poor time management, and inadequate academic writing skills \cite{selemani2018postgraduate}. International students are particularly prone to unintentional plagiarism due to their developing academic literacy \cite{pecorari2024plagiarism}. Addressing these challenges effectively requires a pedagogical response, where institutions provide targeted support to help students develop the necessary skills to adhere to academic conventions \cite{pecorari2024plagiarism}.

Some scholars advocate for educational approaches rather than disciplinary ones. For instance, Gallant (2017) argues that creating a learning-oriented environment that emphasizes mastery over performance can naturally reduce cheating by enhancing students' motivation to learn and developing their metacognitive skills. Similarly, \cite{dinscore2022plagiarism} shows how we can apply an instructional design process to create structured and engaging learning experiences for students to ethically engage with sources. On the other hand, assessment-led solutions have also been proposed. McDonald and Adl \cite{mcdonald2019stopping} recommend low-stakes, formative assessments to combat plagiarism. Hrasky and Kronenberg \cite{hrasky2011curriculum} suggest avoiding tasks that simply require students to collect and present information, advocating instead for essays that integrate theory. However, recent advancements in AI enable individuals to use generative AI models, such as ChatGPT, to summarize sources available on the Internet and write an essay. Because of the capabilities of generative AI models, educators have been struggling to cope with the new wave of student work created with the help of generative AI tools. As questioned by Gao et al. \cite{gao2022comparing}, ensuring that a given report or essay is based on the student’s original work, not just a generative AI output, is a big challenge in educational settings. AI detectors have been developed to deal with the issue, but it is only one side of the coin. The other is changing teaching and assessment in the classroom to help reduce students’ reliance on generative AI tools when working on assessment tasks. One promising way appears to be creating assessments suitable for higher-order cognitive domains of learning. Replacing activities and evaluations that rely on straightforward declarative knowledge with those that call for analysis, synthesis, and evaluation has been suggested (e.g., \cite{van2023generative}). However, there is not enough empirical evidence showing that the impact of generative AI tools on academic dishonesty can be minimized by creating and using assessments that require higher-order cognitive processes. This study aims to contribute to the literature by examining whether the potential impacts of generative AI on academic dishonesty can be minimized with the help of assessment-led solutions.

\section{Literature Review}
\label{sec:headings}
\subsection{Bloom’s taxonomy in education}
Bloom's taxonomy is a hierarchical model that describes cognitive processes, meaning that learning at a higher level depends on skills and knowledge attained at a lower level. The taxonomy's revised version \cite{anderson2001taxonomy} has six major categories (i.e., cognitive processes): Remember, understand, apply, analyze, evaluate, and create. These categories are believed to differ in their complexity, with 'remember' being the least complex and 'create' being the most complex. 
Bloom's taxonomy has been used by schools and teachers in the classroom to specify lesson objectives, design learning activities with different complexities, and prepare assessments (e.g., \cite{nkhoma2016developing}. It has also been used in education to assess engineering curricula (e.g., \cite{dolog2016assessing}), classify course learning outcomes \cite{shaikh2021bloom}, evaluate students' cognitive skills (e.g., \cite{shahzad2024multi}), align learning outcomes with assessment methods (e.g., \cite{gil2016aligning}).

\subsection{Using Bloom’s taxonomy to deter plagiarism}
Bloom’s Taxonomy is widely used to design assessments that not only promote learning but also mitigate academic dishonesty. Assessments that solely require students to engage in lower-order cognitive tasks, like recalling information, are more prone to plagiarism since students can easily find answers via search engines or AI tools like ChatGPT. McGee \cite{mcgee2013supporting} notes that objective tests targeting lower levels of Bloom’s Taxonomy, such as recall or basic comprehension, tend to make cheating more appealing, as students can quickly access answers from external sources. In contrast, open-ended assessments that require higher-order thinking foster deeper learning and are more challenging to fabricate \cite{mcgee2013supporting}. Krsak \cite{krsak2007curbing} similarly argues that short, objective tests, especially when disconnected from prior coursework, increase opportunities for cheating, while tasks that build on prior knowledge reduce it.
Agha et al. \cite{agha2022towards} recommend using case studies or open-ended tasks to assess problem-solving and critical thinking abilities, particularly at higher levels of Bloom’s Taxonomy. These types of assessments compel students to focus on their own skills, making it harder for students to plagiarize. Heckler et al. \cite{heckler2013using} found that assignments encouraging critical thinking lead to less plagiarism than tasks relying on opinions or documentation. Hamilton and Richardson \cite{hamilton2007academic} also argue that real-world problem-solving assessments, particularly those integrating course concepts and skills, are less susceptible to dishonesty. For instance, assignments requiring students to build models using software and consider ethical implications create unique challenges that are hard to replicate dishonestly.
In conclusion, assessments that focus on higher-order cognitive skills, such as application, evaluation, and creation, not only enhance learning but also reduce opportunities for plagiarism. Tasks requiring deep engagement with material or real-world problem-solving help uphold academic integrity.
\section{Present Study}
The current study was guided by the following precept:
Using assessments that require higher-order cognitive processes helps minimize the opportunities to engage in academic dishonesty.
We argue that the impact of generative AI models on academic dishonesty can be minimized by creating and using assessments that require higher-order cognitive processes. We hypothesize that the level of AI plagiarism is higher when tasks designed to foster lower-order thinking skills are used (e.g., remembering and understanding), compared to the AI plagiarism that occurs when tasks requiring the use of higher-order thinking skills are considered in assessments (e.g., analyzing, evaluating, and creating). We also hypothesize that the level of AI plagiarism decreases as task complexity increases, regardless of the technology available to students. To test these hypotheses, we designed three tasks with different complexities using Bloom’s revised taxonomy and created one control group and three treatment groups, each of which was allowed to use only one of the following technologies to complete tasks: e-textbook, Google, and ChatGPT. More specifically, we aimed to address the research questions presented below: 
\begin{itemize}
    \item Does the level of plagiarism (i.e., content similarity) significantly change as task complexity increases? 

    \begin{itemize}
        \item a.	Does this change depend on the technology used to complete tasks (e-textbook, Google, ChatGPT groups vs. the control group)? 
    \end{itemize}
    \item Does the level of AI plagiarism (i.e., AI-generated content) significantly change as task complexity increases? 
    \begin{itemize}
        \item Does this change depend on the technology used to complete tasks (e-textbook, Google, ChatGPT groups vs. the control group)? 
    \end{itemize}
\end{itemize}

\section{Method}
\subsection{Participant}
One hundred fifty-two students enrolled in an information systems department at a private university in Turkey agreed to participate in the study. However, twenty-four students who did not complete all required tasks were removed from the data. Among those who completed all tasks, five students did not follow the instructions accurately. For that reason, they were excluded from the study.  Finally, data from one hundred twenty-three students were used in the study. 
The participants were asked to complete two pretests and two self-efficacy scales measuring their knowledge of data privacy, their knowledge of ChatGPT, search self-efficacy, and search experiences. The average score in the data privacy test was 54.81 (SD = 27.10) out of 100; in the ChatGPT test, it was 71.29 (SD = 9.32) out of 100. The mean score of the search self-efficacy test was 49.52 (SD = 9.22) out of 70. 

\subsection{Research Design}
We used a true-experimental pre-and post-test design with random assignment, involving a control group and three experimental groups: e-textbook, Google, and ChatGPT access \cite{field2002design}. The e-textbook group had access to a PDF version of the course textbook on data privacy, with no other tools allowed. The Google group used only Google search, and the ChatGPT group used only ChatGPT. The control group had no access to these tools. These groups reflect the evolution of media technologies, with e-textbooks representing static content, Google providing dynamic search \cite{jenkins2006new, deuze2007convergence}, and ChatGPT offering generative AI. This design helps examine how these tools impact academic outputs and plagiarism differently, especially compared to more traditional digital resources \cite{manovich2013software}. The experiment was conducted in a controlled lab environment, and continuous monitoring was performed to ensure adherence to the study protocol. Only five students deviated from the procedures but promptly returned to the correct tool when reminded; they were later excluded from the study. This strict supervision helps maintain the internal validity of the findings.
In Task 1, students completed an activity requiring lower-order thinking skills, i.e., remembering and understanding (see Figure 1); in Task 2, they worked on an activity designed to foster medium-order thinking skills, i.e., applying (see Figure 1).  In Task 3, students were asked to propose a solution to the problem they identified, which required the use of higher-order thinking skills, i.e., analyzing, evaluating, and creating (see Figure 1). Participants in each completed the tasks in the following order: Task 1, Task 2, and Task 3. To enhance objectivity, two independent subject experts confirmed the tasks' differentiation according to Bloom's taxonomy, agreeing on the levels for each task. The three tasks mirror the hands-on mini-projects commonly assigned in information systems courses, requiring students to apply theoretical concepts to practical, real-world scenarios.  

\begin{figure}[h!]
    \centering
    \includegraphics[width=\textwidth]{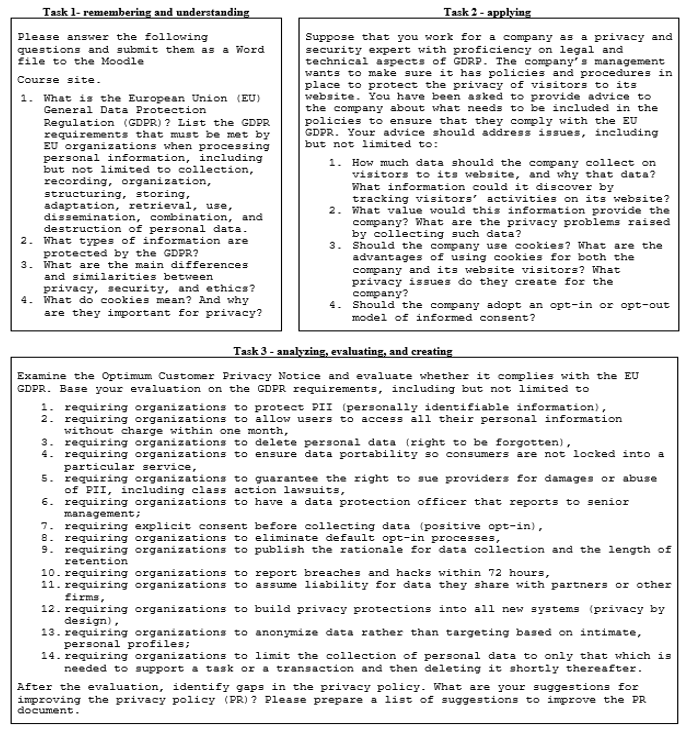} 
    \caption{Tasks used in the study} 
    \label{fig:figure1} 
\end{figure}

\subsection{Group Formation}
We divided participants into four clusters based on their search experience, belief in search skills, knowledge of data privacy, and familiarity with ChatGPT. These factors were chosen due to their influence on information-seeking behavior and technology use, as supported by research. Greater search experience and perceived skills lead to more advanced search strategies \cite{aula2010does, white2009cyberchondria}, while familiarity with tools like ChatGPT affects usage and trust, shaping engagement with technology \cite{kang2016self}. Drawing on these factors, we employed a two-step clustering method to group participants into homogeneous clusters \cite{sarstedt2014concise}. Four clusters were identified. The first cluster (24 participants) had very low composite scores, the second (32 participants) had slightly higher scores, the third (62 participants) had medium scores, and the fourth (35 participants) had the highest scores. To ensure balanced representation, participants from each cluster were evenly distributed across the four experimental groups, maintaining diversity in experience and knowledge. We also ensured gender balance in this process. Figure 2 illustrates the participant clustering and distribution.

\begin{figure}[h!]
    \centering
    \includegraphics[width=\textwidth]{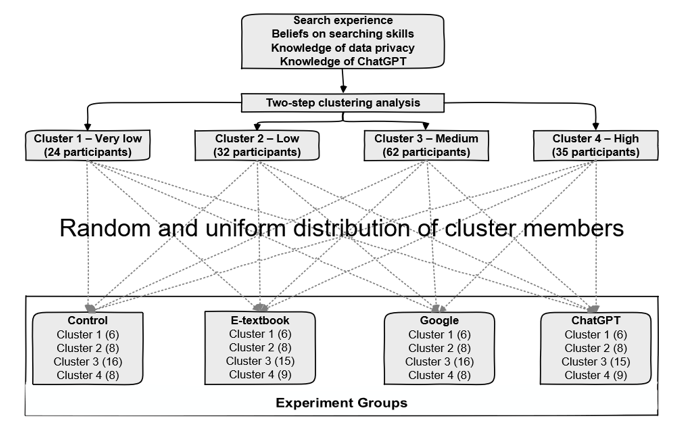} 
    \caption{Group formation} 
    \label{fig:figure2} 
\end{figure}

Table 1 presents the distribution of students by group and gender, with 40 females (32.52\%) and 83 males (67.48\%). The control group included 31 students (25.20\%), while the e-textbook, Google, and ChatGPT groups had 25 (20.33\%), 29 (23.58\%), and 38 (30.8\%) students, respectively. A MANOVA revealed no significant differences before the experiment began, Wilk’s $\lambda$ = .947, F(12, 320.427) = .554, p = .878. Univariate ANOVAs also showed no significant results for search experience, F(3, 124) = .566, p = .638; beliefs about searching skills, F(3, 124) = .086, p = .967; knowledge of data privacy, F(3, 124) = .085, p = .968; and familiarity with ChatGPT, F(3, 124) = 1.239, p = .298. 

\begin{table}[htbp]
\centering
\caption{The frequency distribution of the participants by groups and gender}
\begin{tabular}{lcccccc}
\hline
\multirow{3}{*}{\textbf{Groups}} & \multicolumn{6}{c}{\textbf{Gender}} \\
\cline{2-7}
& \multicolumn{2}{c}{\textbf{Female}} & \multicolumn{2}{c}{\textbf{Male}} & \multicolumn{2}{c}{\textbf{Total}} \\
\cline{2-7}
& \textbf{f} & \textbf{\%} & \textbf{f} & \textbf{\%} & \textbf{f} & \textbf{\%} \\
\hline
Control & 11 & 8.94 & 20 & 16.26 & 31 & 25.20 \\
E-textbook & 10 & 8.13 & 15 & 12.20 & 25 & 20.33 \\
Google & 9 & 7.32 & 20 & 16.26 & 29 & 23.58 \\
ChatGPT & 10 & 8.13 & 28 & 22.76 & 38 & 30.89 \\
\hline
Total & 40 & 32.52 & 83 & 67.48 & 123 & 100.00 \\
\hline
\end{tabular}
\end{table}

\subsection{Procedure}
Before beginning the tasks, the teaching team explained plagiarism in the course context and provided examples. The text given below was included in the instructions for each task. 
\begin{quote}
    In our academic community, plagiarism is defined as the act of using someone else's ideas, work, or expressions without proper acknowledgment. This constitutes academic misconduct, especially when submitting assignments. Examples of plagiarism include: incorporating a specific phrase or sentence from another source without crediting it; directly copying text from books or online materials; closely paraphrasing or translating content from another work; utilizing facts, statistics, graphs, images, or diagrams without acknowledging their origins, and using comments or notes from others, including those derived from lectures or tutorials featuring direct quotes. Even the use of content from assignment writing services or other external aids without proper citation falls under plagiarism. It is essential to employ the correct academic referencing style to clearly indicate the source of any borrowed material in your work.
\end{quote}

For each task, one lecture/demonstration session (50 minutes) and one lab session (100 minutes) were allocated in the course. The lecture/demonstration session took place before the lab session. Lecture/demonstration sessions were used to review relevant course content and concepts and demonstrate how they could be used to accomplish similar tasks. In labs, students were asked to complete the assigned tasks. Upon the completion of each task, submissions were uploaded to Turnitin for plagiarism check via a learning management system. Turnitin recently released its new AI writing detection tool, in addition to its regular similarity scores, which show the percentage of text that was most likely written by an AI writing tool. As a result, each submission underwent a plagiarism check, and Turnitin's percentages served as the basis for generating the data. 

\subsection{Data Analysis and Measures}
After each task, the Turnitin plagiarism detection tool was used to generate similarity and AI plagiarism scores for each submission, as recommended for preventing plagiarism with AI models \cite{cotton2024chatting}. The Similarity Score reflects the percentage of work matching online sources, Turnitin databases, or other papers \cite{turnitin2021understanding}. However, similarity does not always indicate plagiarism, so the teaching team, including the instructor and two teaching assistants, reviewed reports to exclude acceptable similarities like citations and quotes, generating updated scores for the study.
Although AI plagiarism detection technologies often have high false-positive rates \cite{elkhatat2023evaluating}, our study addresses this concern by employing a repeated measures design. We assessed the same group of students on three tasks with varying difficulty levels \cite{field2017discovering}, allowing for internal comparison and reducing error variance, thereby mitigating the impact of false positives \cite{bramwell1992repeated}. While acknowledging that AI plagiarism detectors are not infallible, our focus is on the relative changes in AI plagiarism scores within the same group as they progress through tasks. By comparing students' scores across tasks, we aim to demonstrate patterns in engagement with AI-assisted plagiarism. This approach minimizes the effect of potential false positives on our conclusions, which are based on relative, not absolute, assessments across different levels of task complexity.
The normality of similarity and AI plagiarism scores was examined, and the data obtained in each condition was found to be normally distributed. We applied repeated-measure ANOVA, in which there is a repeatedly measured variable and a group variable with two or more categories \cite{field2017discovering}. IBM SPSS Statistics 27 software was used. As a part of the analysis, the sphericity of both the similarity and AI plagiarism scores were checked \cite{field2017discovering}. The assumption for both variables was not violated: Mauchly's W =.973, 2(2) = 3.178, p =.204, and Mauchly's W =.972, 2(2) = 3.305, p =.192, respectively.

\section{Results}
Table 2 presents descriptive information on Turnitin similarity scores, and Figure 3 illustrates changes in average scores across the group within three tasks. 

\begin{table}[htbp]
\centering
\caption{The descriptive information of similarity scores for three tasks and groups}
\begin{tabular}{lccccc}
\hline
& \textbf{Control} & \textbf{E-textbook} & \textbf{Google} & \textbf{ChatGPT} & \textbf{Total} \\
\hline
\textbf{Task 1} & 4.42 & 49.00 & 51.24 & 42.77 & 36.37 \\
& (12.32) & (30.34) & (30.24) & (15.62) & (29.32) \\
\textbf{Task 2} & 2.58 & 23.68 & 42.76 & 16.11 & 20.52 \\
& (8.97) & (35.66) & (39.09) & (11.07) & (29.42) \\
\textbf{Task 3} & 0.55 & 16.56 & 26.28 & 18.82 & 15.51 \\
& (2.87) & (23.34) & (30.17) & (15.23) & (21.83) \\
\hline
\multicolumn{6}{l}{\small Note. Mean and standard deviation is presented in M (SD) format.} \\
\multicolumn{6}{l}{\small Control n = 31, E-textbook n = 25, Google n = 29 \& ChatGPT n = 37} \\
\hline
\end{tabular}
\end{table}

\begin{figure}[h!]
    \centering
    \includegraphics[width=\textwidth]{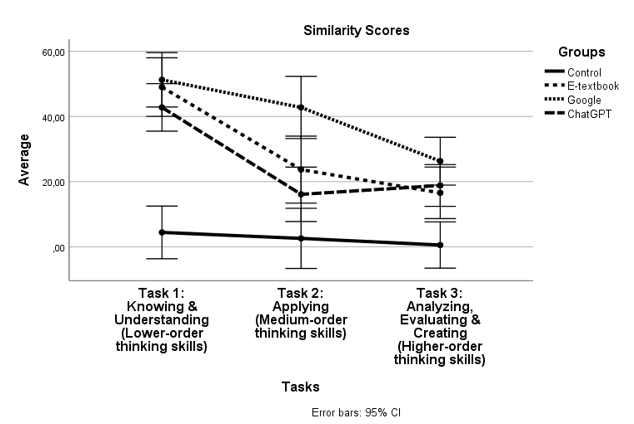} 
    \caption{Similarity scores} 
    \label{fig:figure3} 
\end{figure}

The within-subjects test showed significant results for tasks alone, $F(2, 238) = 46.711$, $p < .01$, partial $\eta^2_\text{p} = .282$, and for the interaction between tasks and groups, $F(6, 238) = 6.030$, $p < .01$, partial $\eta^2_\text{p} = .132$, Cohen's $f = .35$, indicating that 13.2\% of the similarity scores could be explained by this interaction, approaching a large effect size \cite{cohen2013statistical}. The control group’s average was significantly lower than all treatment groups in Task 1 (p < .01), with no significant difference between treatment groups. The Google group had the highest similarity scores across tasks. As task complexity increased, plagiarism rates dropped sharply in treatment groups but only slightly in the control group. The interaction between groups and repeated measures shows that the Control and ChatGPT groups had a consistent performance with lower variability, while the E-textbook and Google groups showed higher variability, particularly in Task 2. 
Table 7 displays descriptive statistics for AI plagiarism percentages, and Figure 5 shows changes in average AI plagiarism percentages across tasks. The within-subjects test yielded significant results for tasks alone, F(2, 238) = 14.475, p < .01, $\eta^2_\text{p}=.108$, and for interaction between tasks and groups, F(6, 238) = 6.620, p < .01, partial $\eta^2_\text{p} = .143$, Cohen's f = .37. The ChatGPT group plagiarized more than all other groups in Task 1 (p < .01), while the Google and ChatGPT groups differed significantly from the control and e-textbook groups in Task 2. Although the differences were smaller in Task 3, the ChatGPT group consistently had the highest AI plagiarism percentages, while the control group had the lowest. As task complexity increased, AI plagiarism decreased in the ChatGPT group, but AI plagiarism remained in the range of 10–20\% in the e-textbook and Google groups, likely due to false positives. The findings suggest a significant decrease in AI plagiarism with increasing task complexity, with the ChatGPT group showing the highest variability and AI plagiarism rates.

\begin{table}[htbp]
\centering
\caption{The descriptive information of AI plagiarism percentages for three tasks by groups}
\begin{tabular}{lccccc}
\hline
& \textbf{Control} & \textbf{E-textbook} & \textbf{Google} & \textbf{ChatGPT} & \textbf{Total} \\
\hline
\textbf{Task 1} & 4.68 & 6.68 & 11.34 & 71.89 & 27.42 \\
& (19.02) & (20.51) & (28.87) & (32.89) & (39.85) \\
\textbf{Task 2} & 0.00 & 9.24 & 21.59 & 64.92 & 27.02 \\
& (0.00) & (27.96) & (36.52) & (37.69) & (39.94) \\
\textbf{Task 3} & 0.00 & 4.68 & 2.28 & 26.74 & 9.75 \\
& (0.00) & (11.39) & (8.53) & (36.74) & (24.17) \\
\hline
\multicolumn{6}{l}{\small Note. Mean and standard deviation is presented in M (SD) format.} \\
\multicolumn{6}{l}{\small Control n = 31, E-textbook n = 25, Google n = 29 \& ChatGPT n = 37} \\
\hline
\end{tabular}
\end{table}

\begin{figure}[h!]
    \centering
    \includegraphics[width=\textwidth]{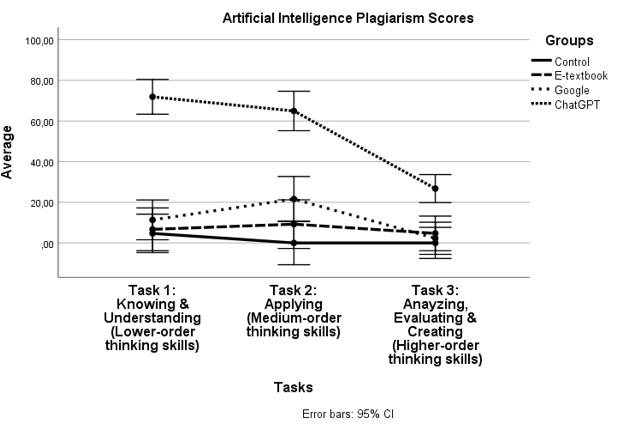} 
    \caption{AI Plagiarism scores} 
    \label{fig:figure4} 
\end{figure}

Finally, we would like to highlight a few key observations from the experiment. In the control group, several participants expressed concerns about the complexity of the tasks and the difficulty of completing them without tools, relying solely on their prior knowledge and experience. In the ChatGPT group, some students found the tool less effective for higher-order tasks, which led to self-doubt about their knowledge. In other groups, participants found the tasks relatively easy, often copying and pasting information, with some attempting to paraphrase and refine it.
\section{Discussion}
With advances in AI, humans began to use AI tools to produce creative outputs in different aspects of their lives. Although generative AI technologies have experienced quick and widespread adoption, their place in education is still up for debate. While some institutions consider generative AI an opportunity to enhance learning and focus on its use in education, others see it as a threat to academic integrity and have been developing rules to restrict students from using it in educational settings \cite{bobrow2023opportunity, pittalwala2023chatgpt}. This study aims to shed light on this debate by examining whether the potential impact of generative AI on plagiarism varies depending on task complexity.

\subsection{AI Plagiarism }
The act of taking someone else’s work and using it without giving them proper credit is a traditional definition of plagiarism. However, with recent advancements in generative AI models and their widespread use in education, AI plagiarism—referring to the use of AI-generated content as one’s original work—has become a growing concern \cite{cotton2024chatting}. This study provides evidence for AI plagiarism and shows it differs from traditional forms of plagiarism. We compared traditional similarity scores with AI plagiarism scores from Turnitin for the same submissions. While there was no significant difference in similarity scores between the three experimental groups, the ChatGPT group's AI plagiarism scores were significantly higher than those of the other groups. These results point out a separation between similarity scores and AI plagiarism scores and suggest using AI plagiarism scores, in addition to similarity scores, to better detect plagiarism.

Additionally, the control group’s average similarity scores were significantly lower than those of all other groups. The higher similarity scores observed in the ChatGPT group compared to the control group can be attributed to the nature of AI-generated text. ChatGPT is trained on a vast dataset that includes a wide array of internet text. While it aims to generate original content, it often produces statistically probable responses based on its training data. This can result in generating phrases, sentences, or even longer passages that closely resemble or match existing texts. Consequently, when submissions from the ChatGPT group are analyzed using plagiarism detection tools like Turnitin, they may yield higher similarity scores due to overlaps with content in these databases. In contrast, participants in the control group relied solely on their knowledge and understanding without access to external sources or tools. Their responses were more individualized and less likely to contain text matching existing online content, resulting in lower similarity scores. This distinction highlights how using AI tools can inadvertently increase the risk of unintentional plagiarism, as AI-generated content may closely mirror existing sources without the user's awareness. This finding underscores the importance of educating students on responsible AI use and the need for instructors to design assessments that mitigate this risk. Educators should guide students on using AI tools ethically, emphasizing the importance of originality and proper citation even when using AI-generated assistance.

This study suggests AI plagiarism scores should be interpreted with caution due to false positives in student submissions. Our findings show that AI plagiarism scores varied from 0–20\% across all groups, even without generative AI use. Thus, we recommend accounting for up to 20\% false positives when interpreting Turnitin's AI plagiarism scores. Elkhatat et al. \cite{elkhatat2023evaluating} raised similar concerns after testing five AI detectors and finding inconsistent performance. Similarly, Chaka \cite{chaka2024reviewing} found that AI and traditional plagiarism detectors remain inconsistent across studies. Chaka's review of 17 articles showed no detection tool, AI-based or traditional, is fully reliable on its own. Therefore, AI detection tools should be combined with traditional plagiarism detectors and human review for better accuracy in identifying AI-generated content. Liu et al. \cite{liu2024great} found that human reviewers, combined with AI detectors, effectively identify AI-generated content, even when paraphrased. Their findings highlight the need for human expertise to complement AI detectors, especially in fields where detection accuracy is critical. They recommend integrating AI detectors into peer review, alongside human reviewers, to better catch AI-generated content.

\subsection{Assessment as a plagiarism prevention strategy}
Our findings show that AI plagiarism decreases as task complexity increases. It peaks when students are asked to recall information or explain concepts, consistent with research showing that assessments based on lower-order thinking skills are more prone to plagiarism \cite{mcgee2013supporting,krsak2007curbing}. Conversely, AI plagiarism is lowest when students must justify decisions, generate new ideas, or propose solutions. This supports the argument that assessments designed to foster higher-order thinking, such as application, evaluation, and creation, reduce academic dishonesty \cite{agha2022towards, heckler2013using}. Postle \cite{postle2009detecting} emphasizes that using assessments aligned with Bloom's taxonomy helps deter plagiarism, while Austin and Brown \cite{austin1999internet} argue that tasks requiring creation make cheating difficult. Given that our findings on AI plagiarism align with what has been suggested in educational assessment, we conclude that assessment practices proposed to deal with plagiarism in the literature remain relevant today in the age of AI.

Traditionally, plagiarism prevention focused on detecting direct copying, but generative AI raises new concerns as it can produce original content without copying \cite{duffy2023chatgpt, roose2023dont}. Our results indicate these concerns are valid, with high AI plagiarism occurring when assessments rely on the replication of authoritative sources but not on their constructive use (for more information on the constructive use of authoritative sources, see Scardamalia \& Bereiter \cite{scardamalia2021knowledge}). It appears that developments in generative AI have the potential to disrupt conventional assessment practices and urge us to be more innovative in assessment design. As our findings suggest, going beyond memorization and recall and designing assessments that engage students in the constructive and productive use of authoritative sources, such as using information in new situations, evaluating the value of information for a specific context, and generating new ideas from existing content, appears to be a more effective prevention strategy that leads to better learning experiences. 

\subsection{Challenges of LLMs in Decision-Making and Opportunities for Assessment Design}
The rapid advancements in generative AI models like ChatGPT-4 and ChatGPT-4o have greatly enhanced their ability to handle complex tasks, particularly in natural language processing, planning, and problem-solving. However, despite these improvements, challenges remain in decision-making, especially in uncertain environments \cite{liu2024dellma, yang2023foundation}. LLMs still struggle with adaptive, multi-step reasoning \cite{dam2024complete}, contextual understanding \cite{zhu2024can}, creativity \cite{franceschelli2024creativity}, and ethical judgment \cite{balas2024exploring}, limiting their reliability in complex decision-making, a key aspect of higher-order cognitive skills.

In our study, which used ChatGPT-3.5, AI-generated plagiarism was more common in lower-order tasks but significantly decreased as task complexity increased. Although newer models like ChatGPT-4o may better handle higher-order tasks, decision-making limitations persist. This presents opportunities for educators to design assessments that challenge the limits of LLMs by focusing on complex decision-making, creativity, and ethical reasoning, reducing reliance on AI-generated content. While LLMs will continue to evolve, assessments emphasizing human judgment, creativity, and decision-making will remain crucial. Instead of excluding AI, these tools can be integrated to support students in developing cognitive skills while ensuring critical thinking isn't fully outsourced to generative models.

\section{Conclusion}
This study examined whether assessments fostering higher-order thinking skills can reduce AI plagiarism. We also explored whether changes in similarity scores and AI plagiarism percentages differ based on the technology used. For this, we designed three tasks of varying complexity using Bloom's updated taxonomy and formed four groups: one control and three treatment groups using either e-textbook, Google, or ChatGPT. The findings reveal a clear distinction between similarity scores and AI plagiarism percentages, supporting the use of both for plagiarism detection. However, given false-positive matches, AI plagiarism percentages should be interpreted cautiously. Overall, our results suggest that assessments promoting higher-order thinking remain a promising approach to prevent AI-driven plagiarism.

\subsection{Limitations and Suggestions for Future Research}
This study was limited by its sample size and the characteristics of participants randomly assigned to the control and treatment groups, as the random assignment was based on only a few key factors. Future studies could include a more diverse sample and broaden the factors used for assignment to improve matching and control. Additionally, all tasks were designed for the data privacy content area. Repeating the study with tasks from other computing areas, such as databases, data structures, computer architecture, and networks, could help assess the transferability of the findings.

The generalizability of these findings may also be limited, as the study was conducted within a specific educational context and content area. Caution should be taken when applying these results to other academic settings. Future research with more diverse participants and across different disciplines would help determine the broader applicability of the findings.

\bibliographystyle{unsrt}  


\bibliography{references}
\end{document}